\documentclass[conference]{IEEEtran}
\usepackage{cite}

\usepackage[cmex10]{amsmath}
\usepackage{array}
\usepackage{mdwmath}
\usepackage{mdwtab}
\usepackage{eqparbox}
\usepackage{graphicx}
\usepackage{color}

\begin{document}
\title{Active optical frequency standards using cold atoms:   
perspectives and challenges}

\author{\IEEEauthorblockN{Georgy A. Kazakov and Thorsten Schumm}
\IEEEauthorblockA{Institute of Atomic and Subatomic Physics, 
Vienna University of Technology\\
 Stadionallee 2, 1020 Vienna, Austria \\
Email: kazakov@thorium.at}}

\maketitle

\begin{abstract}
We consider various approaches to the creation of a high-stability active optical frequency standard, where the atomic ensemble itself produces a highly stable and accurate frequency signal. The short-time frequency stability of such standards may overcome the stability of lasers stabilized to macroscopic cavities which are used as local oscillators in the modern optical frequency standard systems. The main idea is to create a ``superradiant'' laser operating deep in the bad cavity regime, where the decay rate of the cavity field significantly exceeds the decoherence rate of the lasing transition. Two main approaches towards the realization of an active optical frequency standard have been proposed already: the optical lattice laser, and the atomic beam laser. We consider these and some alternative approaches, and  discuss  the  parameters  for atomic ensembles necessary to attain the metrology relevant level of short-time frequency stability, and various effects and main challenges critical for practical implementations.
\end{abstract}


\section{Introduction}
Time and frequency are today the most accurately measured physical quantities by far. Optical clocks using cold particles, i.e. trapped ions and neutral atoms demonstrate unsurpassed performance. The most precise and stable clocks reported up to date have a fractional frequency inaccuracy below $10^{-17}$ \cite{Chou10, Bloom13}, and a short-term stability of about $3.2 \times 10^{-16}\tau^{-1/2}$ \cite{Nicholson12, Hinkley13}. All these high-precision optical clocks are {\em passive frequency standards}, where the frequency of a narrow linewidth laser (local oscillator, LO), prestabilized to an ultrastable optical resonator, is used for interrogation of the quantum discriminator (trapped ion or ensemble of neutral atoms). This resonator undergoes unavoidable thermal \cite{Numata04} and mechanical fluctuations.

Instability of the local oscillator frequency is one of the main factors limiting the short-term stability of modern optical clocks  \cite{Lemke09, Audoin98, Jiang11}. First, fluctuations of the local oscilator frequency entirely determine the overall clock stability on the timescales shorter than one interrogation cycle. Second, down-conversion of the broad-band laser frequency noise also limits the clock stability on longer timescale (Dick effect, see \cite{Audoin98}). Last but not least, instabiliy of the local oscillator frequency limits the interrogation time and hereby prohibits the full exploitation of advantages of narrow transitions available in many optical clock systems.

The best modern reference resonsators provide an Allan deviation of the local oscillator frequency as low as a few $10^{-16}$ in 0.1\,--\,1000~s \cite{Kessler12, Nicholson12}. Further improvement of the short-term stability may be attained not only by means of refinement of the macroscopic reference cavity (which seems to be a challenging task), but also by locking the interrogation laser to another reference, more stable than the macroscopic resonator. A prominent candidate for the role of this reference is an {\em active} optical frequency standard. In the microwave range, similar combinations are used in high-precision timekeeping systems, where the Hydrogen maser (or a set of masers) is used as a fywheel for the primary Caesium standard. In the optical range, there is still no working active systems whose stability could compete with the stability provided by the macroscopic cavities, altough several approaches are already proposed~\cite{Chen05, Yu08, Meiser09} and actively studied. In the present work we overview these and some possible alternative approaches and discuss their perspectives and main challenges appearing on the way towards the practical implementations of metrology-relevant active optical atomic frequency standards.
\section{Modern concepts}

The concept of an active optical frequency standard (or the {\em``clock laser''}) has been recently proposed~\cite{Chen05, Yu08, Meiser09} and studied~\cite{Bohnet12,Bohnet121,Bohnet13,Meiser10, Meiser102,Xu14,Xu13,Bohnet14, Maier14, Chen09, Yang10, Cox14, Zhang12, Kazakov13} by several authors. The main idea is to create a ``superradiant'' laser operating deep in the bad cavity regime, where the decay rate $\kappa/2$ of the cavity field significantly exceeds the decoherence rate $\gamma_{ab}$ of the lasing transition. {In this case, the influence of any fluctuation (thermal or mechanical) of the cavity length on the phase of output radiation is suppressed by a factor of $2\gamma_{ab}/\kappa$~\cite{Kolobov93, Kuppens94} in comparison with conventional lasers}. The two main approaches towards the realization of an active optical frequency standard are: the {\em optical lattice laser}~\cite{Chen05, Meiser09} and the {\em atomic beam laser}~\cite{Yu08}. We will now consider these concepts in more details.

\subsection{Optical lattice laser} 
\begin{figure}
\centering
\includegraphics[width=2.5in]{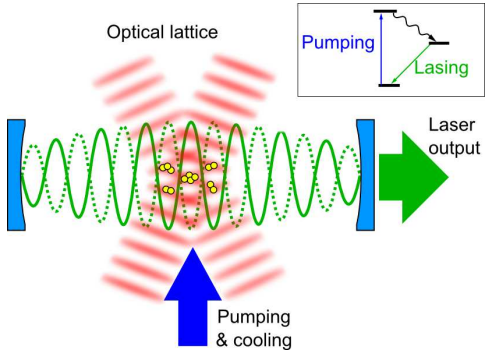}
\caption{Optical lattice laser}
\label{fig_1}
\end{figure}

This approach suggests to use cold atoms with narrow optical transitions (such as  \mbox{${^1\mathrm{S}_0}\leftrightarrow {^3\mathrm{P}_0}$} transitions in divalent atoms), confined to the Lamb-Dicke regime inside an optical lattice potential as a gain medium to build the laser, see Figure \ref{fig_1}. The necessary population inversion can be realized by additional repumping fields, coupling the lower lasing state with some higher levels from which atoms decay to the upper lasing state (3-level laser scheme). Various aspects of optical lattice lasers were studied theoretically in \cite{Meiser09, Meiser10, Meiser102, Maier14, Xu14, Xu13, Bohnet14}. It was shown that in the superradiant regime, the light is to a good approximation coherent. A prototype experiment  using a Raman system to mimic a narrow linewidth optical transition \cite{Bohnet12, Bohnet121, Bohnet13} has demonstrated the general operation principle of such an optical lattice bad cavity laser and applicability of the main theoretical results. 

The threshold conditions for the superradiant (lasing) regime, the power of the output radiation and the linewidth for an ensemble of 2-level atoms coupled to the mode of the optical cavity in the presence of continuous incoherent repumping were considered in \cite{Meiser09}. In this paper, the parameters for quantitative estimations were typical for $^1S_0 \leftrightarrow ^3P_0$ transitions in alkali-earth fermionic isotopes, but not specified. Here we briefly overview the main results and present the specific estimations for this transition in $^{87}$Sr. This element has a nuclear magnetic moment $I=9/2$, spectroscopic transition linewidth $\gamma\,=\,2\,\pi\,\times\,7.6$~mHz \cite{Porsev04} and a transition frequency $\omega_a~\simeq~2~\pi~\times~429$~THz. Suppose that the cavity mode is $\pi$-polarized, and the atoms are pumped into a single spin state $m_F=9/2$. Then, using the Wigner-Eckart theorem, we estimate the dipole transition matrix element as $d_{eg}=\frac{3}{\sqrt{11}}\sqrt{\frac{3 \hbar \gamma c^{3}}{4 \omega_a^3}}\simeq 8 \times 10^{-5} e a_0,$ where $\frac{3}{\sqrt{11}}$ is a Clebsh-Gordan coefficient for the considered transition, $a_0=0.59$~\AA\, the Bohr radius, $e$ the electron charge. Then, taking the cavity length $L=10$~cm and an effective cavity waist radius of 57~$\mu$m, we find the coupling coefficient $g=d_{eg}\sqrt{8\pi\omega_a/(\hbar V_{eff})}\simeq 52$~s$^{-1}$. The cavity finesse $F=10^4$ corresponds to $\kappa=9.41 \times 10^5$~s$^{-1}$.

\begin{figure}
\centering
\includegraphics[width=3.4in]{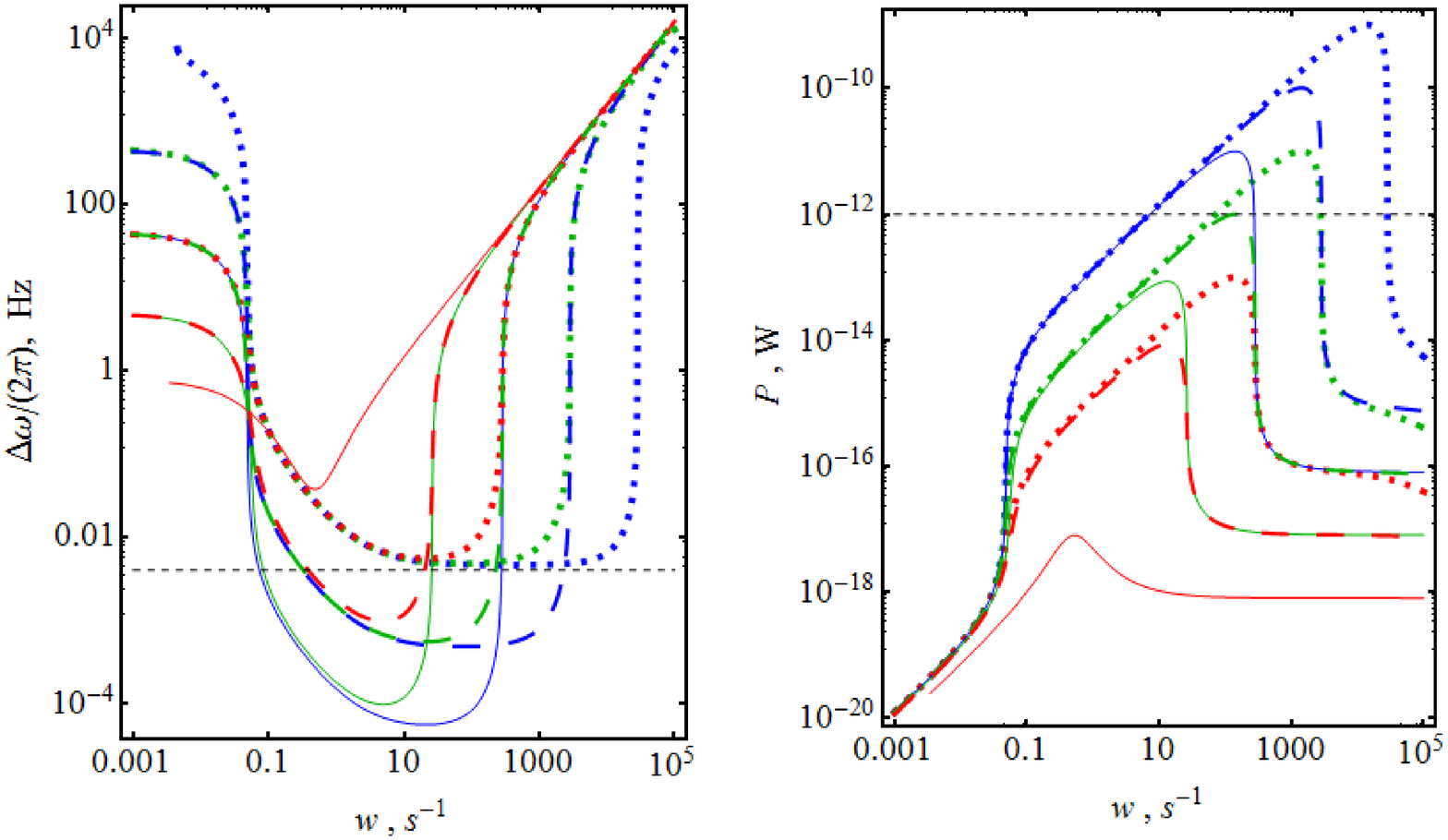}
\caption{Dependence of the cavity field linewidth (left) and the output power (right) on the pumping rate $w$ for 3 different values of finesse $F$ (represented by curve style; solid: $F=10^3$, dashed: $F=10^4$, dotted: $F=10^5$), and for 3 different values of atom number $N$ (represented by color; red: $N=10^4$, green: $N=10^5$, blue: $N=10^6$). Here $\gamma_{ab}=1$~s$^{-1}$, other parameters are specified in the text. Gray dashed lines indicate the target levels for linewidth and output power. }
\label{fig:folpw}
\end{figure}

Our target linewidth is $\Delta \omega=10^{-17}\omega\simeq 2\pi\times 4.3$~mHz. Also, our shot-noise limit estimations shows that 1 pW of output field is enough to lock the phase of the slave laser, providing the relative frequency stability at the level of a few $10^{-17}/\tau$, if the linewidth of this slave laser does not exceed 1 kHz. In Figure \ref{fig:folpw} we represent the dependences of the cavity field linewidth $\Delta \omega/(2\pi)$ and the output power $P$ on the pumping rate $w$ for 3 different values of the cavity finesse, and for 3 different values of the atom number.

We should mention that a number of unresolved problems hampering the creation of an optical lattice laser relevant for metrological applications remains. The lifetime of atoms trapped in the optical lattice is limited. Continuous optical pumping via some intermediate state will heat the atoms hereby reducing the lattice lifetime, and will disturb the lasing transition. First-order Zeeman and vector light shift will also hinder attaining the desirable level of frequency stability, for example, stabilization of the frequency of $^1S_0(m_F=9/2)\leftrightarrow ^3P_0(m_F=9/2)$ transition in $^{87}$Sr at $10^{-17}$ level would require microgauss level of magnetic field stabilization in the timescale of 100~s. We should mention also the collisions between the coupled atoms introducing the density dependent shifts, the Dicke superradiant linewidth enhancement and dipole-dipole interaction between active atoms \cite{Maier14}.

\subsection{Atomic beam laser}

\begin{figure}
\centering
\includegraphics[width=2.5in]{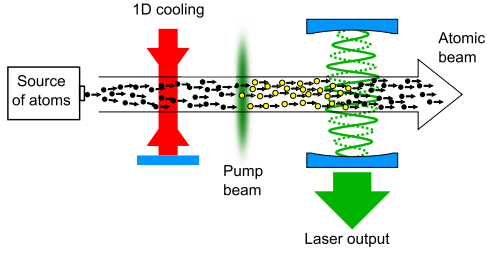}
\caption{Atomic beam laser}
\label{fig:3}
\end{figure}

The second approach is an {\em atomic beam laser}, where a continuous beam of atoms, previously pumped to the upper lasing state, passes through the cavity, see Figure \ref{fig:3}. The theory of laser generation for this system was presented in~\cite{Yu08}, and in \cite{Chen09}, the estimations for several candidate atoms were performed.

Let us summarize the most important results of the theory developed in \cite{Yu08}, taking into account the correspondence \cite{An03} between a realistic Gaussian mode and a sharp cylindric model considered in \cite{Yu08}. Firstly, suppose that all the atoms interact during an equal time $\tau$ with the cavity mode. Then the coupling coefficient $g=C_{CG} \sqrt{\frac{3 \lambda^2 c \gamma}{\pi^2 L w_0^2}}$, where $L$ and $w_0$ are the cavity length and mode waist, and $C_{CG}\equiv C_{F_gm_g1q}^{F_em_e}$ is a Clebsh-Gordan coefficient characterizing the lasing transition.
The steady-state solution is given by the equation
\begin{equation}
\begin{split}
& \sin\frac{\epsilon \tau}{2}=\pm\frac{\epsilon \tau}{2} \sqrt{\frac{\kappa}{R\tau^2 g^2}}, \quad \mathrm{where} \\
& \quad \epsilon=\sqrt{(\omega-\omega_a)^2+\Omega^2},\quad \Omega^2= 4 g^2 n, \quad n=\left\langle \hat{a}^+\hat{a} \right\rangle,
\end{split}
\end{equation}
and $R$ is a total atomic flux (number of atoms which enter into the cavity in one second). The detuning of the radiation from the atomic transition caused by the cavity pulling is:
\begin{equation}
\omega-\omega_a=2\frac{\omega_c-\omega_a}{\kappa \tau}\,\frac{ \epsilon^2 \tau^2 \sin(\epsilon \tau/2)}{\epsilon\tau-\sin(\epsilon\tau)}\, ,
\end{equation}
which yields the bad cavity condition $\kappa \tau \gg 1$ well above the threshold. For the resonant case $\Delta' \ll \Omega$, the threshold condition $R\tau^2 g^2>\kappa$ can be expressed via the cavity finesse $F$ as 
\begin{equation}
R>R_{th}=\frac{\pi^3 v^2}{3 \lambda^2 \gamma F C_{CG}^2} \label{e:6}
\end{equation}
where $v$ is a velocity of the atoms. In the practically interesting case $\gamma \ll \tau^{-1}\ll \kappa$, the laser linewidth becomes $\Delta \omega=g^2/\kappa$.

This threshold seems to be hardly attainable for a free beam of alkali-earth atoms with $^3P_0$ as an upper lasing state. Indeed, for the $^{3}P_0(m_F=9/2)\leftrightarrow^{1}S_0(m_F=9/2)$ transition in $^{87}$Sr atoms, $F=10^5$, and the beam velocity $v=10^2$~cm/s (corresponding to a free fall from 5 cm height), the threshold flux $R_{th}=5.42\times 10^{9}$~s$^{-1}$ is too large for atoms in this upper lasing state. On the other hand, for some other transitions in other atoms, such as $^3P_1\leftrightarrow^1S_0$ in bosonic Ca \cite{Chen09}, the threshold condition seems to be attainable. 

At the same time, the theory developed in~\cite{Yu08} does not take into account the motion of the atoms along the cavity axis while passing through the resonator. This motion will lead to the appearance of a first-order Doppler shift {and related Doppler broadening. The mean first-order Doppler shift of the atomic ensemble will fluctuate together with the mean transversal atomic velocity which depends on the environment, thermal fluctuation of the nozzle etc.} There is another problem connected with the motion of the atoms along the cavity axis. The attainable temperature of the atomic ensemble is about 1 $\mu$K, corresponding to a radial velocity $v_{r}\sim 1$~cm/s. In a standing-wave atomic beam laser, the phase of the intracavity field flips with a spatial period $\lambda/2$ in axial direction. Therefore, the time ${\tau}$ should be less than $\lambda/(2v_r)$, otherwise the atom-field interaction becomes incoherent. At the same time, a bad cavity condition $\kappa \tau \gg 1$ should be fulfilled. Finally, this effect imposes a lower bound on the atomic beam velocity: $v>2 v_r w_0/\lambda$, where $w_0$ is a cavity waist. 

\section{``Combined'' lasing schemes}
As it has been shown in the previous section, realizing a narrow and stable laser based on a beam of ballistic atoms seems to be unfeasible. However, this scheme rests on the promising idea to spatially separate the pumping and lasing processes (as in masers), and it is free of the principal limitations in the time domain connected with the final lifetime of the atoms in the optical lattice. We propose several schemes that in some sense can be considered as ``hybrids'' of the optical lattice and atomic beam lasers. 

To keep the atoms in the Lamb-Dicke regime along the cavity axis, and to overcome the restrictions described in the end of the previous section, we propose to use an {\em auxiliary blue-detuned magic optical lattice} confining these atoms in the cavity waist of the clock laser. Atoms can be prepared in the upper lasing state outside the cavity, circumventing perturbations due to ac Stark shifts. 
\begin{figure}
\centering
\includegraphics[width=2.7in]{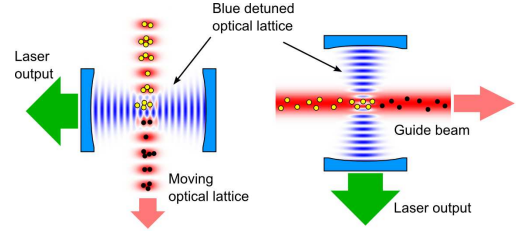}
\caption{Lasing schemes with continuous preparation of inverted atoms: with vertical optical conveyor (left), and with horizontal guide beam (right).}
\label{fig:f4}
\end{figure}
Ballistic atoms in the Earth gravity field seem to be too energetic and not sufficiently focused to be manipulated by such blue-detuned optical lattice potential. Therefore, it becomes necessary to use slow atoms guided by an {optical conveyor}, or by a {horizontal red-detuned magic wavelength laser beam}, see Figure~\ref{fig:f4}. In the first case, a Bessel beam could be exploited to provide a transversal gradient of laser intensity sufficient to support the atoms against gravity, like in \cite{Schmid06}. A vertical optical conveyor formed by frequency-shifted counterpropagating beams would allow to use a much weaker intensity of the red-detuned field because of a much higher intensity gradient along the vertical axis. On the other hand, the horizontal blue-detuned optical lattice should be much deeper than in the scheme with horizontal guide beam, to avoid tunneling and the emergence of a band-like spectrum \cite{Lemonde05}. Certain technical difficulties may arise due to the required intensities.

\begin{figure}
\centering
\includegraphics[width=2.7in]{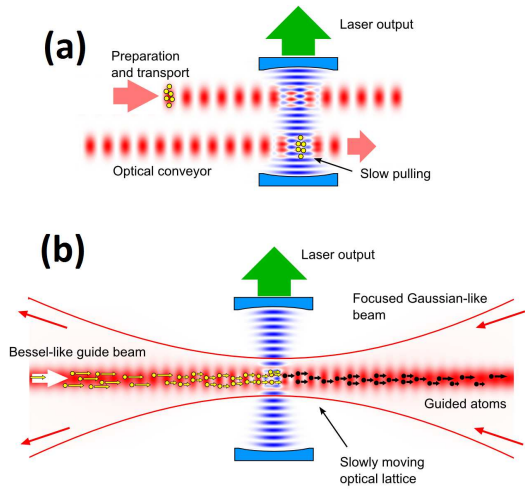}
\caption{Lasing schemes with irregularly moved atoms.}
\label{fig:f5}
\end{figure}

Optical conveyors allow to pull atoms through the cavity waist slowly, hereby increasing the time $\tau$ during which atoms interact with the laser mode. This method both improves the bad cavity condition $\kappa\tau \gg 1$, and decreases the threshold atomic flux $R_{th}$ in comparison with a free beam scheme, see equation (\ref{e:6}). On the other hand, slow speed and limited lifetime of atoms in the optical conveyor potential may not allow to place the laser cavity far enough from the zone where the atoms are cooled and loaded. To overcome this contradiction, one can {\em transport} the atoms from the preparation zone {\em quickly}, and then {\em pull} them through the cavity waist {\em slowly}. One of such methods, namely the {sequential coupling of atomic ensembles} was proposed by us \cite{Kazakov13}, see Figure \ref{fig:f5}a. In this method, atomic ensembles preparied in the upper lasing state outside the cavity can be introduced by turn into the cavity. Maintaining of the optical phase cam be attained by proper choice of timing, i.e. the ``fresh'' ensemble should be introduced into the cavity when the ``old'' one still emit some radiation into the cavity mode, hereby fresh ensemble start to emit photons mainly via stimulation emission.

Another interesting possibility is to use a moving optical lattice formed by a { horizontal Bessel beam and a counter-propagating frequency shifted Gaussian beam} tighly focused to the clock cavity waist, see Figure \ref{fig:f5}b. In this case, the moving lattice exists only in the lasing zone, and it can provide controllable and slow motion of atoms through the cavity waist, whereas far from the cavity waist the atoms can move freely and more quickly along the axis of the beam.


\section{Conclusion}
In the present paper, we overviewed two existing approaches to the creation of active optical frequency standards, namely atomic beam laser, and the optical lattice laser. Parameters of the optical resonator and atomic ensemble, necessary for the generation of laser radiation suitable for stabilization of the frequency of some slave laser at the level of $10^{-17}$ {seems to be attainable within the optical lattice laser approach}. Certain problems connected with the limited lifetime of the atomic ensemble in the optical lattice potential, heating of the atoms by the repumping field and with the control on the Zeeman and vector light shifts should be solved. We have proposed several ``combined'' schemes, where the atoms will be prepared in the upper lasing state outside the cavities and then transported into the cavity by a guide beam or optical conveyors, and the blue-detuned optical lattice will prevent motion of atoms along the cavity axis hereby preventing the decoherence and first order Doppler effect.

\section*{Acknowledgment}

The authors would like to thank Helmut Ritch, Murray Holland and especially Florian Schreck for the helpful discussions. The work was supported by  the Austrian Science Fund (FWF) through project I-1602.

\end{document}